\documentclass[journal]{IEEEtran}
\usepackage{cite,url,subfigure,epsfig,graphicx}
\usepackage{tipa}
\usepackage{makecell}
\usepackage{multirow}
\usepackage{bbm}
\usepackage{mathrsfs}
\usepackage{array}
\usepackage{amsfonts}
\usepackage{cite,url,subfigure,epsfig,graphicx}
\usepackage{amssymb,amsmath}
\usepackage{indentfirst}
\usepackage{pifont} 
\usepackage{cases}
\IEEEoverridecommandlockouts
\makeatletter
\def\ps@headings{
\def\@oddhead{\mbox{}\scriptsize\rightmark \hfil \thepage}
\def\@evenhead{\scriptsize\thepage \hfil \leftmark\mbox{}}
\def\@oddfoot{}
\def\@evenfoot{}}
\makeatother 

\usepackage{times,verbatim,amsfonts,amsmath,color}
\begin{document}
\title{Mobile Wireless Rechargeable UAV Networks: Challenges and Solutions}

\author{
\IEEEauthorblockN{Yuntao~Wang$^{1}$, Zhou~Su$^{1}$, Ning~Zhang$^{2}$, and Ruidong~Li$^{3}$}\\
\IEEEauthorblockA{
$^{1}$School of Cyber Science and Engineering, Xi'an Jiaotong University, China\\
$^{2}$Department of Electrical and Computer Engineering, University of Windsor, Canada\\
$^{3}$Institute of Science and Engineering, Kanazawa University, Japan\\
$^{*}$Corresponding author: Zhou~Su (zhousu@ieee.org)
}}

\maketitle

\begin{abstract}
Unmanned aerial vehicles (UAVs) can help facilitate cost-effective and flexible service provisioning in future smart cities.
Nevertheless, UAV applications generally suffer severe flight time limitations due to constrained onboard battery capacity, causing a necessity of frequent battery recharging or replacement when performing persistent missions.
Utilizing wireless mobile chargers, such as vehicles with wireless charging equipment for on-demand self-recharging has been envisioned as a promising solution to address this issue. 
In this article, we present a comprehensive study of \underline{v}ehicle-assisted \underline{w}ireless rechargeable \underline{U}AV \underline{n}etworks (VWUNs) to promote on-demand, secure, and efficient UAV recharging services.
Specifically, we first discuss the opportunities and challenges of deploying VWUNs and review state-of-the-art solutions in this field.
We then propose a secure and privacy-preserving VWUN framework for UAVs and ground vehicles based on differential privacy (DP).
Within this framework, an online double auction mechanism is developed for optimal charging scheduling, and a two-phase DP algorithm is devised to preserve the sensitive bidding and energy trading information of participants. Experimental results demonstrate that the proposed framework can effectively enhance charging efficiency and security. Finally, we outline promising directions for future research in this emerging field.
\end{abstract}

\IEEEpeerreviewmaketitle

\section{Introduction}
\IEEEPARstart{B}{enefiting} from the advances in hardware and wireless communication techniques, recent years have witnessed the proliferation of unmanned aerial vehicles (UAVs) in numerous civilian and commercial applications such as crowd surveillance, agriculture mapping, and disaster assessment. It is anticipated that the global UAV market value will reach USD 58.4 billion by 2026. 
Due to the prominent features of low cost, high maneuverability, and rapid deployment, UAVs can be immediately dispatched to environmental-harsh areas such as mountains, oceans, and disaster areas to carry out missions in an on-demand manner \cite{UAVref1}. Moreover, with the mounted rich onboard sensors, UAVs can cooperatively perceive the information in a given area and transmit data back to the ground station for intelligent decisions.
For instance, in disaster-stricken areas, UAVs can capture images and detect survivors to assist disaster recovery and rescue. Meanwhile, a swarm of UAVs can quickly establish the emergency communication network and act as aerial relays to facilitate the dissemination of disaster-critical information when terrestrial network infrastructures are not available \cite{UAVref2}.

Owing to stringent space and weight constraints, commercial UAVs, especially small UAVs such as quadrotors, generally have inherent battery limitations to support their long-term operations, which restricts their flight time and flight range.
For instance, micro UAVs (i.e., length less than 100~cm and weight less than 2~kg) powered by Lithium Polymer batteries, which constitute almost 90\% of the micro UAV market, usually have a maximum flight endurance of 90 minutes \cite{chargingReview}. 
Nevertheless, the real-time compute-intensive missions, e.g., video streaming and image processing, in complicated UAV applications may consume a considerable amount of energy and is detrimental to the battery lifetime.
Notably, in complex missions requiring extensive battery endurance, the battery constraint issue can not be addressed by only optimizing UAV's energy management.
Therefore, effective battery recharging is urgently needed for the practical deployment of UAV applications.

\begin{figure*}[t]\setlength{\abovecaptionskip}{-0.0cm}
\centering
  \includegraphics[width=15.8cm]{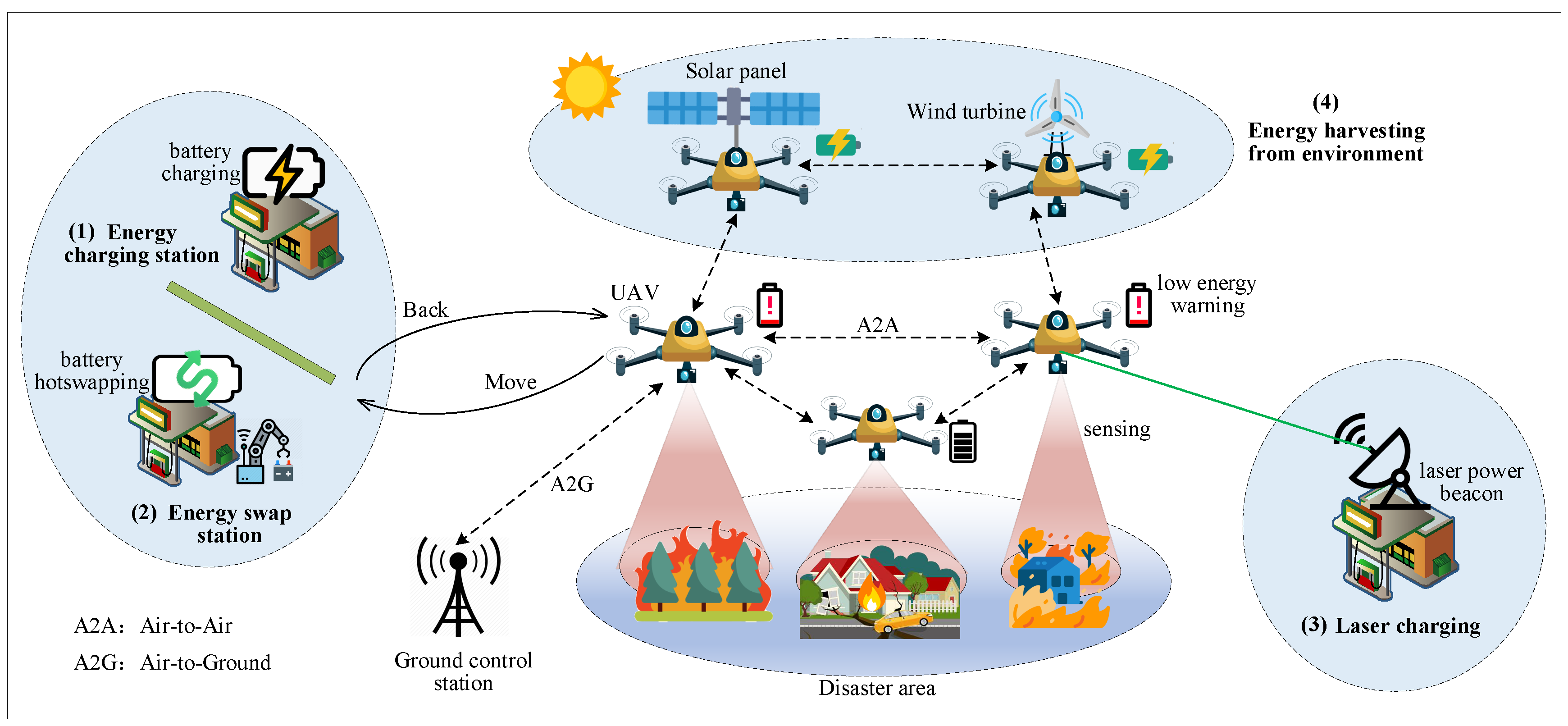}
  \caption{Overview of battery recharging methods for inflight UAVs when performing persistent missions: (1) flying to the energy charging station (CS) for recharging; (2) flying to the energy swap station for battery hotswapping; (3) using the laser power beacon (PB) for wirelessly battery charging; (4) equipping auxiliary renewable energy (RE) generators to harvest energy from the environment.}\label{fig:1}
\end{figure*}

In the literature, various battery recharging methods have been presented to recharge UAV's depleted battery when executing outdoor tasks \cite{Added2021WPT,CSCharging,SwapCharging,LaserCharging,RECharging,StaticWPT,UGVCharging2}, which can be classified into two main types: (i) obtaining energy from a wired/wireless electrical source (e.g., static charging stations, battery hotswapping, and laser charging), and (ii) directly harvesting energy from the surrounding environment (e.g., equipping solar panels). 
In the former approach, solutions based on static wired chargers usually need human interventions in battery recharging/replacing and lack feasibility for UAVs' charging operations.
In the latter approach, the outfitted photovoltaics (PV) arrays on UAV's wings may add additional weight, size, and operational difficulty to safely land dedicated locations. Moreover, this approach highly depends on weather conditions.
With the mature of the promising wireless power transfer (WPT) technology, wireless mobile chargers (WMCs) can address this issue by utilizing unmanned ground vehicles (UGVs) equipped with wireless charging facilities to realize on-demand and fully automatic UAV battery recharging \cite{UGVCharging2}.
However, as the size of charging pads built on vehicle platforms needs to be comparably smaller than that of fixed chargers, it is challenging to support multiple UAVs' charging operations due to the limited energy capacity.
Consequently, the cooperation between UAVs and UGVs is of necessity, and well-designed incentives are needed to motivate selfish UGVs' cooperation.
In addition, as UAVs and UGVs are mutually distrustful entities, there exist inherent security and privacy concerns in the open energy trading environment.
For instance, their locations, driving/flying routes, and energy states can be inferred by insiders and external adversaries.
In summary, to realize secure and efficient energy transfer in \underline{v}ehicle-assisted \underline{w}ireless rechargeable \underline{U}AV \underline{n}etworks (VWUNs), the following three main challenges require to be resolved: 1) fast and on-demand UAV recharging, 2) efficient scheduling of UAVs and UGVs, and 3) privacy protection of participants.

In this article, we present a comprehensive study of VWUNs to facilitate on-demand, secure, and efficient UAV recharging services.
Specifically, the motivations and challenges of deploying VWUNs in practical scenarios are presented, followed by a review of existing advanced solutions in this field.
Then, we exploit differential privacy (DP) and auction theory to build a secure and efficient recharging framework for distrustful UAVs and UGVs.
Under this framework, we further develop a privacy-preserving online double auction (PPODA) mechanism to jointly optimize charging management and protect participants' private energy trading and bidding information in a distributed manner.
Simulation experiments are conducted to evaluate the proposed framework. At last, several future research directions are outlined in VWUNs.

The remainder of this article is organized as follows. Section II introduces the background, main challenges, and existing solutions of VWUNs. Section III presents the design of the proposed framework for VWUNs, and Section IV shows a case study to evaluate the proposed framework. Section V outlines the open research directions, and Section VI closes the article with conclusions.

\section{Background, Challenges, and Existing Solutions of wireless-powered UAV Networks} 

\subsection{Rechargeable UAV Networks}

In commercial UAV applications, the inherent battery endurance constraints restrict UAVs from performing persistent missions. The design of rechargeable UAV networks has become a critical issue for practical UAV deployments. Fig.~\ref{fig:1} shows four typical UAV battery recharging methods as below. 

1) \textbf{Flying to the fixed charging station (CS) \cite{CSCharging}.} As shown in the left part of Fig.~\ref{fig:1}, the UAV can fly to a fixed CS to recharge its depleted battery and move back to continue missions. This method has been widely adopted in practice to charge multiple UAVs simultaneously. However, as the chargers are fixed, it lacks feasibility for UAVs to obtain energy conveniently. Moreover, it can incur high round-trip energy costs in the flight to/from the CS. 

2) \textbf{Battery hotswapping \cite{SwapCharging}.} Nowadays, most UAVs are designed with pluggable battery packs to enable fast battery replacement. As shown in the left part of Fig.~\ref{fig:1}, during hotswapping process, UAV's depleted battery can be replaced by a fully charged one in the battery swap station, with the advantage of avoiding the long waiting time for battery charging. The drawback is that it usually depends on the human labor to detach the depleted battery and insert a replenished one into the UAV.

3) \textbf{Laser-beam inflight recharging \cite{LaserCharging}.} As shown in the right part of Fig.~\ref{fig:1}, a laser power beacon (PB) can be deployed on the rooftop of buildings and transmit laser beams to the UAV that enters its aerial power link coverage. Then, UAV's embedded laser receiver converts the laser into electrical power. This enables UAVs to stay inflight without landing for battery recharging/replacing. However, it incurs high deployment costs and requires a line-of-sight path from the laser to the UAV which is hard to guarantee in urban cities with buildings of varying heights.

4) \textbf{Energy harvesting from environment \cite{RECharging}.} As shown in the upper part of Fig.~\ref{fig:1}, UAVs can also use renewable energy (RE) generators such as solar cells and wind turbines as auxiliary power sources in executing persistent tasks (e.g., more than one day). However, this approach is highly weather-dependent, and the additional weight and size of RE generators (e.g., the outfitted PV arrays on UAV's wings) can add operational difficulty for safely landing in dedicated sites.

\begin{table*}[]
\caption{Comparison of inflight UAV Recharging Methods in Various Applications}\label{UAVRechargingMethods}
\resizebox{\textwidth}{!}{%
\begin{tabular}{|c|c|c|c|c|c|c|c|}
\hline
\textbf{Name}     & \textbf{Type}  & \textbf{Pros}                                                                             & \textbf{Cons}                                                                                                       & \textbf{\begin{tabular}[c]{@{}c@{}}Human \\intervention level\end{tabular}} & \textbf{\begin{tabular}[c]{@{}c@{}}Energy \\ efficiency\end{tabular}} & \textbf{\begin{tabular}[c]{@{}c@{}}Additional \\ flight\end{tabular}} & \textbf{Ref.} \\ \hline
Cable charging    & Wired          & Low deployment cost                                                                       & Very limited working area                                                                                              & High                                                                         & High                                                                 & No                                                                    & ---              \\ \hline
CS-based charging & Wired/Wireless & Support multi-UAV charging                                                                & \begin{tabular}[c]{@{}c@{}}High round-trip energy cost\\ Low charging feasibility\end{tabular}                      & Medium                                                                       & High                                                                  & Long
& \cite{CSCharging}             \\ \hline
Battery hotswapping  & Swap           & Support multi-UAV charging                                                                & \begin{tabular}[c]{@{}c@{}}High round-trip energy cost\\ Issues in autonomous swapping\end{tabular}                      & Medium                                                                       & Very high                                                                 & Long
& \cite{SwapCharging}               \\ \hline
Laser PB charging  & Wireless       & \begin{tabular}[c]{@{}c@{}}No need to land\\ No additional flight\end{tabular}           & \begin{tabular}[c]{@{}c@{}}High deployment cost\\ Need complete UAV motion information\end{tabular}                                & Medium                                                                       & Low                                                                   & No
& \cite{LaserCharging}              \\ \hline
RE-based charging & \begin{tabular}[c]{@{}c@{}}Harvesting energy\\ from environment\end{tabular}           & \begin{tabular}[c]{@{}c@{}}No need to land\\ No additional flight\end{tabular}           & \begin{tabular}[c]{@{}c@{}}Weather-depend\\ Limited harvested energy\\Need additional weight and size\end{tabular}         & No                                                                           & Medium                                                               & No
& \cite{RECharging}              \\ \hline
Stationary WPT    & Wireless       & \begin{tabular}[c]{@{}c@{}}High charging feasibility\\ No human intervention\end{tabular} & Need additional flight                                                                                              & No                                                                           & Medium                                                               & Medium                                                                & \cite{StaticWPT}              \\ \hline
UGV-assisted WPT  & Wireless       & \begin{tabular}[c]{@{}c@{}}On-demand self-recharging\\ No human intervention\end{tabular}      & \begin{tabular}[c]{@{}c@{}}Complex route/resource/landing\\ scheduling\end{tabular}                                                                               & No                                                                           & Medium                                                                & Short
& \cite{UGVCharging}              \\ \hline
UAV-assisted WPT  & Wireless       & \begin{tabular}[c]{@{}c@{}}On-demand self-recharging\\ No need to land\end{tabular}      & \begin{tabular}[c]{@{}c@{}}Hard to operate autonomously\\ Easy for aerial collision\end{tabular} & No                                                                           & Medium                                                                & Short                                                                 &  \cite{UAVCharging}             \\ \hline
\end{tabular}
}
\end{table*}

\begin{figure}[t]\setlength{\abovecaptionskip}{-0.0cm}
\centering
  \includegraphics[width=9.2cm]{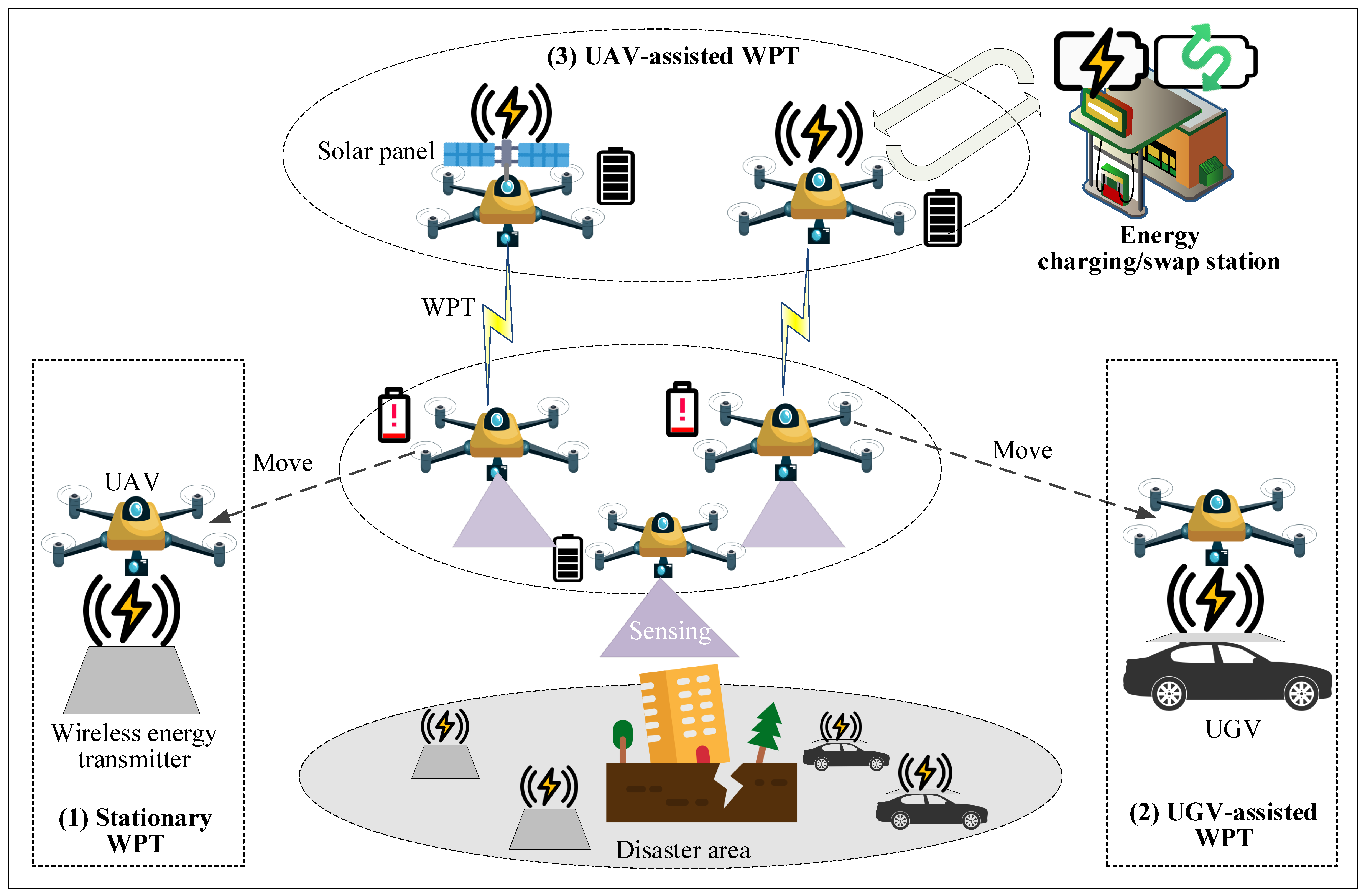} \vspace{-0.35cm}
  \caption{Illustration of WPT-based battery recharging methods for inflight UAVs: (1) stationary WPT; (2) UGV-assisted WPT; (3) UAV-assisted WPT.}\label{fig:2}
\end{figure}

Recently, with the mature of WPT technology, wireless-powered UAVs are proposed as an alternative approach to extend the battery lifetime of UAVs. Fig.~\ref{fig:2} illustrates three scenarios of WPT-based UAV battery recharging as below.

1) \textbf{Stationary WPT \cite{StaticWPT}.} As shown in the left part of Fig.~\ref{fig:2}, a UAV equipped with a WPT device can fly to a nearby charger with a wireless energy transmitter (ET) to wirelessly and automatically recharge its depleted battery. The wireless ETs can be installed at cell towers, power poles, rooftops, etc.

2) \textbf{UGV-assisted WPT \cite{UGVCharging}.} As shown in the right part of Fig.~\ref{fig:2}, by installing wireless ETs on the roofs of UGVs, UAVs can receive on-demand self-recharging services due to the controllable mobility of both UAVs and UGVs.

3) \textbf{UAV-assisted WPT \cite{UAVCharging}.} As shown in the upper part of Fig.~\ref{fig:2}, in a UAV swarm, part of UAVs with WPT ETs, called energy hosts (EHs), can serve as energy sources to recharge other UAVs in the air. The energy of EH UAVs can come from solar/wind energy or ground supplements for fast replenishing. One of the main challenges is to operate autonomously and safely with collision avoidance.

Table \ref{UAVRechargingMethods} summaries the detailed comparison of different UAV recharging methods in various applications.

\subsection{UGV-Assisted Wireless Rechargeable UAV Networks}
\begin{figure*}[t]\setlength{\abovecaptionskip}{-0.0cm}
\centering
  \includegraphics[width=17cm]{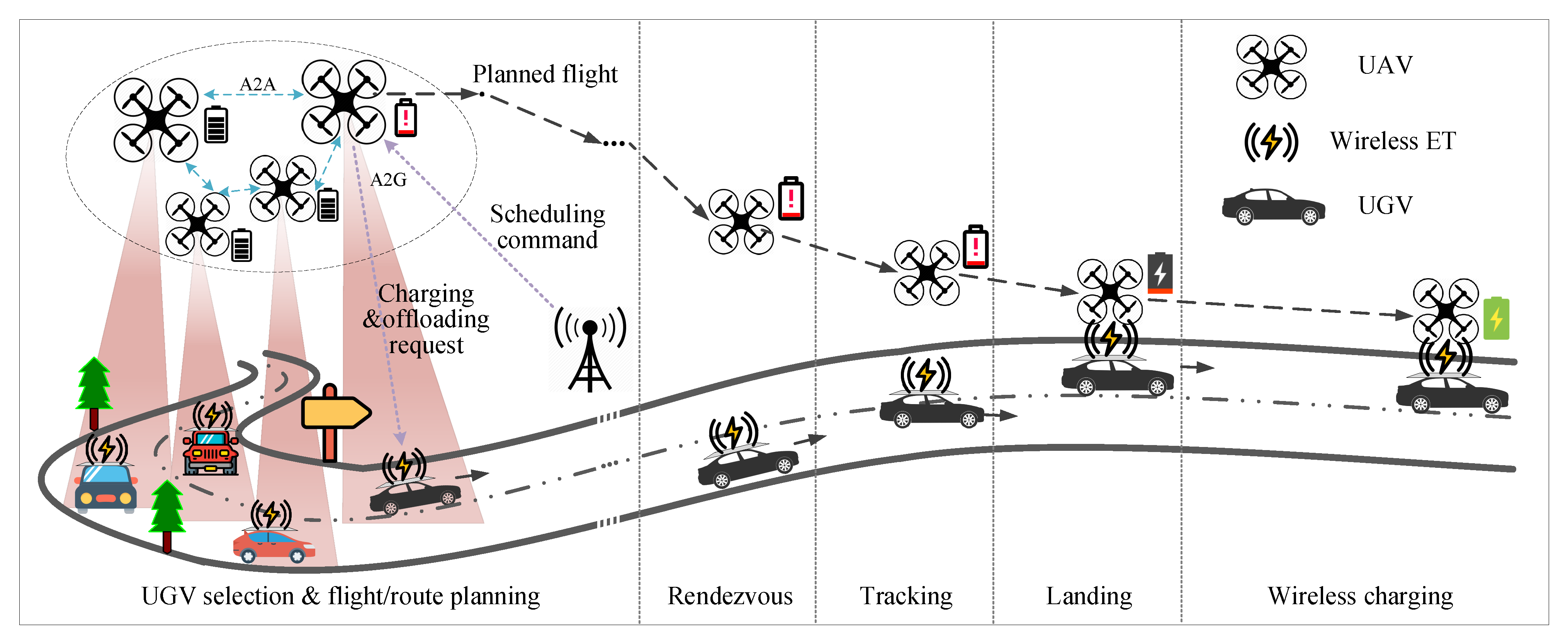} 
  \caption{Process of multi-UGV-assisted self-recharging for multiple UAVs under WPT.}\label{fig:3}
\end{figure*}

The fixed wired chargers and battery hotswapping for UAV recharging usually require human labors and have low charging feasibility, which consequently impacts UAV's flight time and operation efficiency.
Wireless mobile chargers (WMCs) \cite{UGVCharging2} are proposed as an alternative approach by leveraging UGVs with WPT facilities to enable on-demand and fully automatic battery recharging for UAVs.
Fig.~\ref{fig:3} illustrates the detailed procedure of multi-UGV-assisted self-recharging for multiple UAVs, which consists of the following six phases. 

Specifically, when the battery state-of-charge (SoC) drops to a certain level, the UAV sends its charging request to nearby UGVs with sufficient energy resources via air-to-ground (A2G) links.
In the \emph{UGV selection phase}, the UAV selects an appropriate UGV and negotiates the rendezvous to receive recharging service according to the remaining battery SoC and their locations. 
The UAV also constantly receives the scheduling command from the ground control station to avoid collision and maintain air traffic.
After that, in the \emph{route planning phase}, the UAV plans its flight and the UGV plans its driving route. In the next \emph{rendezvous phase}, both of them move to the mutually agreed rendezvous.
Once the UGV or the UAV gets close to the rendezvous, it enters the automatic tracking mode. During the \emph{tracking phase}, upon the UAV is captured by the object tracking camera equipped on the UGV, the UGV automatically adjusts its positions beneath the UAV. Then, the UAV enters the automatic landing mode. When the UAV lands on the landing pad (i.e., the wireless charging pad) on the roof of the UGV, the \emph{landing phase} finishes.
In the last \emph{wireless charging phase}, the UAV continuously receives energy from the wireless ET of the UGV until attaining a desirable SoC level. 

\subsection{Challenges of Energy Transfer Over VWUNs}
With the increasing demand for secure and efficient self-recharging services in UAV applications, new challenges arise in energy transfer over VWUNs.

\textbf{Fast and on-demand UAV recharging: }
As shown in Fig.~\ref{fig:3}, VWUN offers a potential solution to provide on-demand and automatic UAV recharging services by employing cooperative UGVs with WPT devices. Nevertheless, due to the space limits of vehicle roofs, the charging pads built on UGV platforms may suffer capacity limitations to support multiple UAVs' charging operations. In addition, as multiple UAVs and UGVs with distinct energy desires may coexist within the task area, their cooperation and competition in energy services need to be considered. Besides, current works on UAV recharging mainly focus on the one-to-one charging model and many-to-one charging model between UAVs and WMCs, where the many-to-many charging pattern is neglected.
Hence, fast and on-demand recharging mechanisms with high feasibility and automation for UAVs should be designed.

\textbf{Route and charging scheduling of UAVs and UGVs: }In VWUNs, UAVs with low battery SoC need to plan their 3D flights according to their SoC levels and locations of nearby UGVs and UAVs to minimize the energy consumption for battery replenishing while avoiding collisions. Meanwhile, for UGVs, they need to plan their 2D driving routes by considering the geographical restriction and UAV's task area to maximize their revenues. Based on the mutually negotiated rendezvous, UGV's charging capacity, and UAV's charging desires, efficient charging strategies are required to determine the optimal charging duration.
In addition, due to the high mobility of UAVs and UGVs and the correspondingly fast-changing network topology in VWUNs, a distributed mechanism for route and energy scheduling is more suitable for UAVs and UGVs to make real-time decisions.

\textbf{Privacy protection of participants: }
As the VWUNs are open-accessed to facilitate recharging services, UAVs may be vulnerable to a variety of attacks, e.g., jamming attacks and DDoS attacks, conducted by adversaries. Besides, as UAVs and UGVs are mutually distrustful, potential security vulnerabilities and privacy violations may occur in such an untrusted environment. For instance, strategic UGVs may overclaim their charging costs to receive higher compensations from UAVs, which is unfair for honest participants. Moreover, malicious entities may launch inference attacks to steal participants' privacy such as locations, energy costs, and traded energy volumes.
In addition, there exists a dilemma in ensuring both strategy proofness and privacy preservation, as the strategy proofness makes UGVs reveal true charging cost types while privacy targets at preventing others from learning these private information.
Therefore, it is necessary to defend against strategic UGVs while enforcing privacy preservation for honest participants. 

\subsection{State-of-the-Art Solutions}
\begin{figure*}[!ht]\setlength{\abovecaptionskip}{-0.0cm}
\centering
  \includegraphics[width=18.cm]{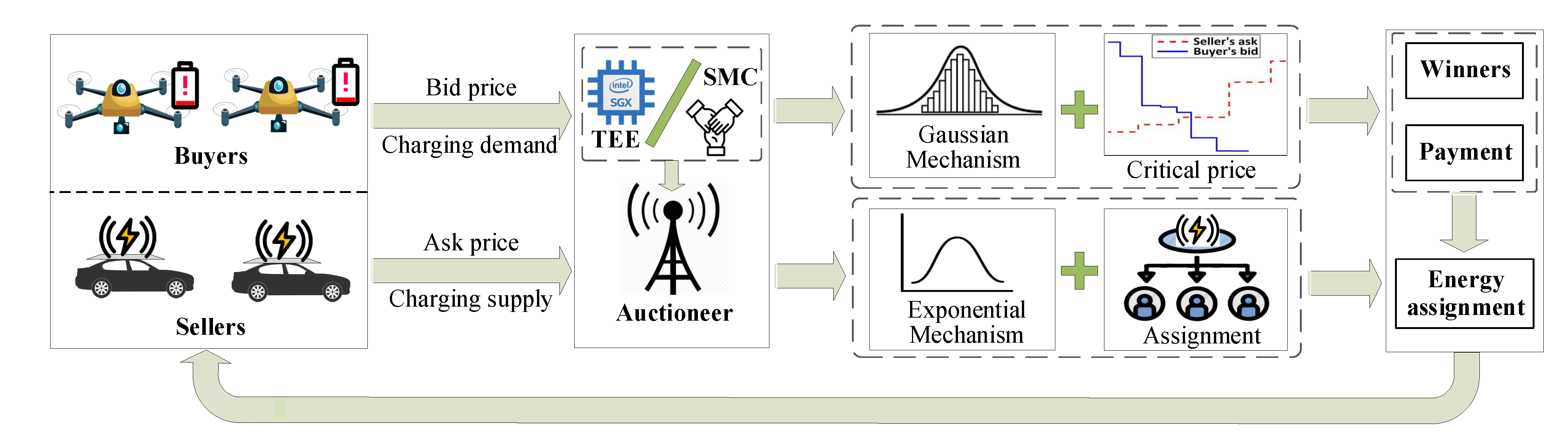} 
  \caption{A privacy-preserving double auction framework for UAV recharging.}\label{fig:4}
\end{figure*}

Of late, many works have been done to address secure and efficient UAV recharging from the aspect of route planning, charging scheduling, and privacy protection.

\textbf{Route planning: } By leveraging edge computing and computer vision techniques, Chen \emph{et al}. \cite{UAVref1} propose an online autonomous flight method for connected UAVs to plan collision-free trajectories from the constructed 3D map.
Ribeiro \emph{et al}. \cite{UGVCharging2} study a collaborative routing mechanism for UAVs and WMCs, in which a variant of vehicle routing problem is formulated for UAV recharging scheduling in rescue missions.
Fu \emph{et al}. \cite{Fu2021IoTj} design a Q-learning based distributed route planning algorithm for UAVs in data collection, where UAVs hover over the static wireless charger to replenish energy while collecting sensory data at each spot.

\textbf{Charging scheduling: }By discretizing UAV's flight paths into spatiotemporal segments, Li \emph{et al}. \cite{StaticWPT} present an efficient charging time scheduling algorithm for wireless chargers to minimize energy consumption. Shin \emph{et al}. \cite{UGVCharging} present a reverse auction-based charging mechanism for multiple UAVs, where each UAV bids for available charging time slots of WMCs in a distributed fashion. 
Xu \emph{et al}. \cite{UAVCharging} devise a Q-learning approach for dynamic and distributed energy scheduling of UAV swarms, where part of UAVs act as the WMCs to offer wireless energy for other UAVs.

\textbf{Privacy protection: }Existing privacy protection mechanisms are mainly built on the trusted computing, DP, federated learning, and cryptographic tools (e.g., homomorphic encryption (HE) and secure multi-party computation (SMC)). Liu \emph{et al}. \cite{UAVPrivacy1} combine HE and trusted computing techniques to preserve clients' privacy when UAV collects and processes their private data. 
Wang \emph{et al}. \cite{UAVPrivacy2} introduce DP-enhanced privacy-preserving sensory data sharing for collaborative UAVs under the federated learning paradigm. 
However, the HE and SMC approaches may consume a considerable amount of resources for battery-limited UAVs in practical scenarios. 

\section{Privacy-Preserving UAV Recharging Framework Under WPT}

\subsection{Design Overview}
Existing UAV recharging approaches mainly focus on the one-to-one charging model \cite{SwapCharging,LaserCharging} or many-to-one charging model \cite{UGVCharging,UAVCharging} between UAVs and UGVs, and few of them consider the joint defense of strategic participants and privacy preservation during energy scheduling. 
As shown in Fig.~\ref{fig:4}, we take an econometric approach by formulating the double-side charging scheduling process between multiple UAVs and UGVs as an online double auction with the target to maximize the overall payoff of involved UAVs and UGVs while ensuring strategy proofness and privacy preservation.
We focus on the scenario in which both UAVs and UGVs cooperatively perform the persistent missions in the task area, such as disaster search and rescue in disaster sites.
Three kinds of entities are involved in the proposed framework: auctioneer (i.e., the ground control station), buyers (i.e., UAVs), and sellers (i.e., UGVs).
The auctioneer is supposed to be trusted, or otherwise a trustworthy auctioneer with SMC or trusted hardware enclave is adopted.
In the proposed online auction, UAVs and UGVs are allowed to bid anytime and the bid collection process ends until reaching a maximum waiting time. 
Moreover, the many-to-many charging pattern is taken into account, where UAVs compete for battery recharging while UGVs compete to provide wireless energy.
The bidding information of UAV (or UGV) includes bidder identity, bidding price (or ask price) per unit energy, desirable energy demand (or supply), and minimum energy demand (or supply).
The bidding/ask price is set based on its valuation of per unit energy demand/supply such as the recharging urgency and energy service cost. 

\subsection{Threat Model}
\textbf{Strategic UAVs and UGVs: }Strategic participants may misreport the valuations and manipulate their bidding/ask prices to gain high economic benefits in the auction. For example, a strategic UAV may report an untruthful bidding price less than its valuation to gain more revenue if it wins. As a result, the enthusiasm of honest participants can be damped and the fairness of the auction market can be undermined.

\textbf{Bidding privacy leakage: }Adversaries may carry out attacks such as inference attacks to deduce users' private bidding information such as the sensitive user valuations and demand/supply volumes via multiple queries from the public auction outcomes. By learning these information, users' energy SoC, locations, and economic situation can be leaked.

\subsection{Privacy-Preserving Online Double Auction (PPODA)}
The goal of our PPODA mechanism is to enforce strategy proofness and privacy preservation for UAVs and UGVs in VWUNs. The DP method is employed to ensure that adversaries cannot infer the true input of any user from the perturbed output with strong probability. 
Here, we leverage the Gaussian mechanism and exponential mechanism to protect users' bidding privacy including valuations and energy volumes.
As illustrated in Fig.~\ref{fig:4}, PPODA consists of the following phases. 

\emph{Phase 1: Online bid collection. }The auctioneer announces the privacy parameters in DP and collects bids from bidders (i.e., UAVs and UGVs) within the maximum waiting time. 

\emph{Phase 2: Privacy-preserving winner and payment determination. }After receiving users' bids, the auctioneer sorts sellers' ask prices in non-decreasing order and buyers' bids in non-increasing order. Then, the auctioneer determines the seller's clearing price (SCP) and buyer's clearing price (BCP) based on the intersection point of the demand and supply curves. To prevent the leakage of bidders' valuations from the public clearing prices, as shown in Fig.~\ref{fig:4}, the Gaussian mechanism is exploited by adding the random Gaussian noise sampled from the Gaussian distribution to each bidder's valuation. 
The expectation of the Gaussian distribution is zero, and the covariance is related to the privacy budget and sensitivity.
The sensitivity is set as the maximum of the difference between two valuations of bidders.

Based on the perturbed valuation input, the clearing prices can be computed. The identities of UAVs whose bid price is no less than BCP is added into the potential winning buyer set, and the identities of UGVs whose bid price is no larger than SCP is added into the potential winning seller set. 
As the number of winning sellers should be equal to that of winning buyers, we randomly determine \emph{w} winning sellers and \emph{w} winning buyers from these two sets, where \emph{w} is the minimum number of potential winning buyers/sellers. To ensure strategy proofness, the widely used uniform pricing method is adopted to prevent bidders from manipulating the auction, where the SCP and BCP are assigned as the payments to winning UGVs and winning UAVs, respectively. Moreover, the preservation of bidders' valuations in this phase is ensured by the Gaussian mechanism.

\emph{Phase 3: Privacy-preserving energy assignment. }
The exponential mechanism is employed to preserve UAVs' energy demand volumes, as depicted in Fig.~\ref{fig:4}. For simplicity, we suppose that the energy supply of UGVs can satisfy any UAV's desirable charging volumes.
Specifically, we build an integer set \emph{E} to denote all available energy assignments for a UAV. 
For a UAV, the first element in \emph{E} is its minimum energy demand while the last element is its desirable energy demand. Every element in \emph{E} is one larger than its previous element.
For every element in the integer set, its quality function \emph{Q} is calculated as this element divided by the UAV's desirable energy volume.
For each energy assignment, its adoption probability can be derived based on the exponential mechanism, where the numerator is the exponential function of the product of the privacy budget and the corresponding quality function, and the denominator is the normalization coefficient.
Then, the probability distribution can be computed for all possible energy assignments of the UAV. Under the probability distribution, an energy assignment is randomly opted by the UAV as its energy transaction volume. Therefore, the preservation of bidders' energy trading volume is enforced by the exponential mechanism.

\emph{Phase 4: Release auction outcome. }The auctioneer announces the winners and their payments, as well as their energy assignments to all participants. Then, it updates the system states and prepares for the next round of auction. For UAVs that fail to match a desired UGV, they can fly to a nearby fixed charging/swap station for recharging.

\section{Case Study}

\begin{figure}[t]
\fbox{\centering
\label{fig:6-2}
\includegraphics[width=7.5cm]{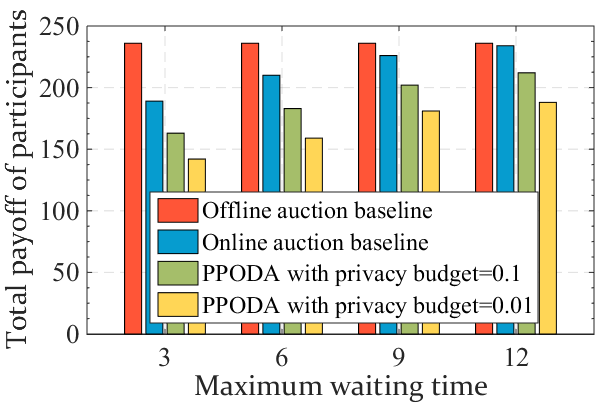}}
\caption{Total payoff of participants vs. maximum waiting time under four schemes.}
\label{fig:5}\vspace{-3mm}
\end{figure}
A simulation scenario on a two-lane road, whose width and length are 20m and 3km respectively, is considered. In this case study, 30 UAVs are uniformly distributed over the road (the distance between each pair of neighboring UAVs is 100m), and 40 UGVs are randomly deployed with minimum safe distance of 30m. UAVs fly at a constant altitude of 5m with the maximum flying speed 10~m/s. The speed of UGVs varies from 20~km/h to 60~km/h. The battery capacity of UAV is 17~Wh. UAVs' demand volumes are randomly selected from [5,10]~Wh. The wireless energy conversion efficiency is 80\%. The valuations of bidders (i.e., UAVs and UGVs) are uniformly distributed from 0 to 15~cents/kWh. Two types of bidders are considered: strategic and honest.

We employ the offline auction scheme and online auction scheme as baselines. The offline auction scheme solves the double auction problem in an offline manner without waiting time constraints.
Fig.~\ref{fig:5} illustrates the overall payoff of all involved participants under four schemes given different values of maximum waiting time.
As seen in Fig.~\ref{fig:5}, with higher waiting time, the gap between online auction and offline auction becomes smaller. The reason is that more bidders can be included in the online auction when the waiting time is large, thereby causing an improved overall payoff. Moreover, the proposed PPODA attains a smaller payoff of involved participants than two baselines, and the gap becomes larger with the decrease of the privacy budget.
It can be explained as follows. On one hand, additional random processes are introduced into the winner\&payment determination and energy assignment to enforce DP at the expense of decreased payoff of participants. On the other hand, a lower privacy budget indicates stronger privacy protection, as well as a larger deviation between the raw data and the perturbed data.

\section{Future Research Directions}
\subsection{Intelligent and Cooperative Routing}
Cooperative route planning in VWUNs is essential for efficient rendezvous generation and energy minimization for both UAVs and UGVs. Due to the highly dynamic network topology and potential geographic restrictions, how to intelligently manage UGVs' routes and UAVs' flights in a distributed, real-time, and collision-free manner becomes a critical challenge. Multi-agent deep reinforcement learning will be a potential solution for intelligent and distributed routing scheduling. A major research issue is to accelerate the learning speed and enable efficient knowledge transfer.

\subsection{Joint Scheduling Under Mixed Recharging Mode}
Future UAVs may be powered by various kinds of energy sources such as solar cell, fuel cell, and battery to prolong battery endurance in persistent tasks. Under such a hybrid power supply, it is necessary to efficiently manage UAVs' recharging behaviors. Moreover, UAVs can offload part of computing missions to nearby UGVs to further save the limited battery. Therefore, the joint scheduling of computing, communication, and hybrid energy resource between UAVs and UGVs while considering their cooperation and competition becomes a crucial concern.

\subsection{Trust-Free Security Provisioning}
Existing works on VWUN mainly rely on a trusted third party or trusted hardware, whose malfunction or outage can cause a system failure. Therefore, it is important to offer a trust-free approach to build trust and prevent fraud for mutually distrustful UAVs and UGVs in the untrusted VWUN environment. Blockchain technology provides a potential solution to ensure transparent resource trading and security provision under public audit. A critical issue is to design a lightweight and robust blockchain system to tolerate potential network fragmentation in VWUNs due to UAVs' high mobility and constrained resources.

\section{Conclusion}
In this article, we have investigated the secure and efficient wireless-powered UAV networks. Several representative UAV recharging architectures have been reviewed, and the opportunities, critical challenges, and existing solutions of deploying VWUNs for on-demand and privacy-preserving UAV recharging have been discussed. Then, we have presented a differential private framework for UAV recharging under VWUNs, in which two DP-based algorithms are integrated with the online double auction-based charging scheduling mechanism to ensure user payoff optimization with bid privacy protection. A case study has been given to validate the effectiveness of the proposed framework. At last, research issues essential for VWUNs are discussed.

\section*{Acknowledgement}\label{sec:Acknowledge}
This work was supported in part by NSFC (nos. U20A20175, U1808207), the Fundamental Research Funds for the Central Universities, and JSPS KAKENHI Grant Number 19H04105.

\end{document}